# Mass and angular momentum assessement of the neutron star XTEJ 1807-294 using X-ray triplets

Radostina P. Tasheva[1a)], Ivan Zh. Stefanov[1b)]

[1]*Technical University-Sofia, Faculty of Applied Mathematics and Informatics, Department of Applied Physics, 8, Kliment Ohridski Blvd., 1000 Sofia, Bulgaria*

[a)] rpt@tu-sofia.bg
[b)] izhivkov@tu-sofia.bg

**Abstract.** The X-ray power-density spectrum of the accreting millisecond pulsar XTEJ 1807-294 contains several triplets of simultaneously observed quasi-periodic oscillations (QPOs): a lower and a upper kHz QPOs and a low-frequency QPO. This fact allows us to assess the mass and the specific angular momentum of the central neutron star. We also apply an error propagation formula to obtain interval estimates of the parameters of the star. The calculations are repeated with several different models for the kHz QPOs. The degree of consistency of the bounds obtained with the different triplets may provide information for the viability of the used models.

## INTRODUCTION

Beeing remaint of massive stars makes neutron stars objects with both strong gravitational and magnetic field that hardly allows light to leave. One way to receive information about these objects is to investigate the accretion disc consisting of remnants of the supernova explosion that surround them. After its depletion which happens in relatively short time the neutron star left with no source of energy dies. If the neutron stars is the heavier companion of a binary system though when the other star reaches the stage of the red giant it statrs “feeding” the neutron starts and the accretion disc increases in size and become hotter and faster. This revives the neutron star itself through the falling accretion matter making it millisecond pulsar. The greatest amont of the energy emited by the accretion disc is X-ray emission and due to the periodicity of its motion quasiperiodic oscilations are observed in the form of repeating peaks[1].

The peak values of the X-ray power density spectrum correspond to the frequencies of the QPOs , which are related to the mass *M* and specific angular momentum *a* of the neutron star [2] .The low frequent osccilations (LF) are shifting and do not exceed 100 Hz. High frequency QPOs tend to appear in pairs – lower - $\nu_L$ and upper $\nu_U$ , very often related as
$\nu_U : \nu_L$ =3:2.

The aim of our investigation is using data from the X-ray power-density spectrum of the accreting millisecond pulsar XTEJ 1807-294 to to assess the mass and the specific angular momentum of the central neutron star. As to do this we use several triplets of simultaneously observed quasi-periodic oscillations. The estimates for the mass and the specific angular momentum are given by the solution of a system of nonlinear algebraic equations. To evaluate the errors of the obtained estimates we use error propagation formulas.

## THE MILLISRCOND PULSAR XTEJ 1807 -294

The X-ray pulsar XTEJ 1807-294 is discovered in 2003 by means of Rossy X-ray Timing Explorer. Linares et.al.[3] report the presence of both low and paired high components in each of the seven groups of data, from B до H given in Table 1.These oscilations are believed to arise simultaneously and on the same orbits in the accretion disc. In order to assess their value modeling with multi-lorenzian function is used.

**TABLE 1.** Observational frequencies for the millisecond pulsar XTEJ 1807-294.

| Group | $\nu_{LF}$(Hz) | $\nu_L$(Hz) | $\nu_U$(Hz) |
|---|---|---|---|
| B | 14.45±0.17 | 163±23.0 | 354.0±10.0 |
| C | 17.4±0.4 | 191.0±9.0 | 375.0±2.0 |
| D | 20.2±0.4 | 202.0±11.0 | 395.0±3.0 |
| E | 26.1±1.6 | 238.0±25.0 | 449.0±9.0 |
| F | 27.6±0.5 | 259.0±16.0 | 465.0±2.0 |
| G | 32.1±1.5 | 273.0±19.0 | 492.0±5.0 |
| H | 40.3±1.1 | 370.0±18.0 | 565.0±5.0 |

## QPO FREQUENCY MODELING

In order to explain the peak values in the X-ray spectrum the model of a hot particle emission is adopted.The particle originates from the companion star and is assumed to have orbitational frequency which is a combination of the three geodesic frequencies- orbital $\nu_\Phi$, radial $\nu_r$ and vertical $\nu_\theta$ according to Aliev et al[2]. The low frequency is ascribed to the Lense-Thirring precession: $\nu_{LF} = \nu_{nod} = \nu_\Phi - \nu_\Theta$ due to the different axis orbits of the cetral object and accretion disc. The lower $\nu_L$ and the upper $\nu_U$ frequencies depend on the model that predicts the creation of the X-ray peaks.

The “relativistic precession” model RP is firstly proposed by Stella & Vietri [4] and after this modified by Torok et al. [5] into the version RP1 and by Bursa into the version RP2. The differences between the versions are due to the upper frequency model chosen (see Table 2). The test particle is believed to move in a quasieliptic orbit with a radius bigger then Innermost stable circular orbit (ISCO).

The total precession model TP and TP1 are firstly introduced by Stuchlık et.al. [7] and exploit the total precession period[7] – the time for which the test particle will be at the closest distance from the star. This alows them to propose a new combination for the lower frequency $\nu_L = \nu_\theta - \nu_r$, and the upper frequency is $\nu_\theta$ for the TP model and $\nu_\Phi$ for the TP1 model.

The warped disc model WD[8] considers an excitation mechanism involving internal viscid and adiabatic perturbation of a deformed accretion disc. This leads to a horizontal resonance responcible for the QPOs created because of warping due to the tilt of the accretion dics axis with respect to the star axis and the consecutive frequency superposition.

**TABLE 2.** Models for the upper $\nu_U$ and lower $\nu_L$ frequiencies

| Model | $\nu_L$ | $\nu_U$ | Autors |
|---|---|---|---|
| RP | $\nu_\Phi - \nu_r$ | $\nu_\Phi$ | Stella & Vietri [4] |
| RP1 | $\nu_\Phi - \nu_r$ | $\nu_\theta$ | Bursa [6] |
| RP2 | $\nu_\Phi - \nu_r$ | $2\nu_\Phi - \nu_\theta$ | Torok et al. [5] |
| TP | $\nu_\theta - \nu_r$ | $\nu_\theta$ | Stuchlık et.al. [7] |
| TP1 | $\nu_\theta - \nu_r$ | $\nu_\Phi$ | Stuchlık et.al. [7] |
| WD | $2(\nu_\Phi - \nu_r)$ | $2\nu_\Phi - \nu_r$ | Kato [8] |
| TD | $\nu_\Phi$ | $\nu_\Phi+\nu_r$ | Cadez et al. [9] |

The tidal disruption model TD [8] assumes strong tidal forces that disrupt and distort the test particle.It then orbits around the neutron star in a almost eliptical orbit.

## FREQUENCY EQUATIONS AND THE ERROR PROPAGATION METHOD

The free parameters of the models, namely the specific angular momentum *a*, the mass of the cenrtal object *M* and the specific radius *r* of the orbit where the corresponding frequencies originate, are implicit functions of the observed frequencies:

$$a = a\left(\nu_{LF_0}, \nu_{L_0}, \nu_{U_0}\right) \quad (1)$$
$$M = M\left(\nu_{LF_0}, \nu_{L_0}, \nu_{U_0}\right)$$
$$r = r\left(\nu_{LF_0}, \nu_{L_0}, \nu_{U_0}\right)$$

through the following system of non-linear algebraic equations

$$f_1 \equiv \nu_{LF}(M,a,r) - \nu_{LF_0} = 0, \quad (2)$$
$$f_2 \equiv \nu_L(M,a,r) - \nu_{L_0} = 0,$$
$$f_3 \equiv \nu_U(M,a,r) - \nu_{U_0} = 0.$$

The observed values of the frequencies are designated by the subscript “0”.
In order to assess the errors of the parameters we apply the following error propagation formulas

$$\sigma_a^2 = \left(\frac{\partial a}{\partial \nu_{U_0}}\right)^2 \sigma_{\nu_{U_0}}^2 + \left(\frac{\partial a}{\partial \nu_{L_0}}\right)^2 \sigma_{\nu_{L_0}}^2 + \left(\frac{\partial a}{\partial \nu_{LF_0}}\right)^2 \sigma_{\nu_{LF_0}}^2, \quad (3)$$

$$\sigma_M^2 = \left(\frac{\partial M}{\partial \nu_{U_0}}\right)^2 \sigma_{\nu_{U_0}}^2 + \left(\frac{\partial M}{\partial \nu_{L_0}}\right)^2 \sigma_{\nu_{L_0}}^2 + \left(\frac{\partial M}{\partial \nu_{LF_0}}\right)^2 \sigma_{\nu_{LF_0}}^2.$$

The uncertainties of the observed frequencies, namely $\sigma_{\nu_{U_0}}, \sigma_{\nu_{L_0}}, \sigma_{\nu_{LF_0}}$ are taken from Table 1.

Here the rules for implicit function differentiation are used for the calculation of the partial derivatives. For more details we refer the reader to [11].

The best estimate for the mass $M_{av}$ as well as its stantandart error $\sigma^2_{av}$ are given by the weighed average over the three independent measurements:

$$M_{av} = \frac{\sum M_i / \sigma_i^2}{\sum 1/\sigma_i^2} \qquad \sigma_{av}^2 = \frac{1}{\sum 1/\sigma_i^2} \quad (6)$$

For the specific angular momentum we use similar formulas.
Here $M_{av}$ is the weighed mass and $\sigma^2_{av}$ is the standard deviation. For the specific angular momentum we use similar formulas.

## MASS AND SPECIFIC ANGULAR MOMENTUM OF THE MILLISECOND PULSAR XTEJ 1807 -294

The mass and the angular momentum are calculated for each of the time intervals from B to H. The weighed average mass and specific angular momentum for all intervals according to the TP,TP1 TD and WD models are presented in Table 3 and according to the RP, RP1 and RP2 models in Table 4. The mass is calculated in solar mases and the angular momentum has no dimension.

**TABLE 3.** Weighed mass and angular momentum of XTEJ 1807-294. for TP,TP1, TD and WD models

| model | $M_{av}$ , solar masses | $a_{av}$, dimensionless |
|---|---|---|
| TP | $4.74 \pm 0.01$ | $0.41 \pm 0.01$ |
| TP1 | $5.30 \pm 0.08$ | $0.41 \pm 0.01$ |
| TD | $6.78 \pm 0.13$ | $0.78 \pm 0.10$ |
| WD | $6.05 \pm 0.23$ | $0.75 \pm 0.02$ |

**TABLE 4.** Weighed mass and angular momentum of XTEJ 1807-294. for relativistic precession models

| model | $M_{av}$ , solar masses | $a_{av}$, dimensionless |
|---|---|---|
| RP | $4.99 \pm 0.09$ | $0.44 \pm 0.01$ |
| RP1 | $4.44 \pm 0.08$ | $0.44 \pm 0.01$ |
| RP2 | $5.64 \pm 0.10$ | $0.45 \pm 0.04$ |

As it is shown in Table 3, values for the mass for TP and TP1 models are lower than for TD and WD models - below 5.30 solar masses for TP and TP1 models and above 6 solar masses for TD and WD models. The assessed specific angular momentum is also higher for TD and WD models – above 0.7.

The results for the relativistic precession models shown in Table 4 do not differ significantly from the results for TP and TP1 models – masses below 5.7 solar masses and angular momentum around 0.4. The highest mass exibits RP2 model.

As a whole the estimates for the the mass are not appropriate for a neutron star and the specific angular momentum is not competable with the Kerr metrics.

In [13] Tasheva and Stefanov have investigated XTEJ 1807-294 and comletely different results were obtained - $M_{opt}=(11.2\pm0.1)$ solar masses – approximately twice as much compared to the present work. They used only lower $\nu_L$ and upper $\nu_U$ high frequencies i.e. the low frequency $\nu_{LF}$ is not involved in the calculations. The optimal values for M and a are required for which the model describes all the seven pairs of observational frequencies that are presummmed to arise simultaneously[12]. Therefore it is possible that including low frequency $\nu_{LF}$ might lead to improvement of the obtained results.

In comperison, the calculations for the neutron star 4U 1728-34 by Tasheva and Stefanov[11] done by usage of the frequency triplets show values for the mass appropriate for a neutron star and agreeable specific angular momentum.

## CONCLUSION

We use the data for the peak values of the frequencies exhibited by the millisecond pulsar XTEJ 1807-294 corresponding to seven time intervals in order to assess the mass and the angular momentum of the object. The data are processed by Linares et.al.[3] and include low ($\nu_{LF}$) and the two high frequencies - lower ($\nu_L$) and upper ($\nu_U$) of the peak values found in the power density X-ray spectra of the neutron star.In order to do this we solved a system of three non-linear algebraic equation. The models utilized - TP,TP1,TD,WD as well as the three relativistic precession models use combinations of fundamental geodesic frequencies with expressions corresponding to Aliev et al (3).

All of the masses obtained are bigger than those appropriate for a neutron star – 3 solar masses. The angular momentum values received are around 0.4 for all models except for WD and TD models.

## ACKNOWLEDGMENT

The authors would like to thank the Research and Development Sector at the Technical University-Sofia for the financial support.